\documentclass[conference]{IEEEtran}
\ifCLASSINFOpdf
\else
\fi
\usepackage{amsmath}
\usepackage{amssymb}
\usepackage{cite}
\usepackage{url}
\usepackage{float}
\usepackage{subfig}
\usepackage{graphicx}
\usepackage{subfig}
\usepackage{footnote}
\newcommand{\cov}{\textrm{Cov}}

\def\thick#1{\hbox{\rlap{$#1$}\kern0.25pt\rlap{$#1$}\kern0.25pt$#1$}}

\def\by{{\bf y}}

\begin{document}
%
\title{\LARGE \bf Real-Time Estimation of the Distribution of Brake Response Times for an Individual Driver 
 Using Vehicular Ad Hoc Network}

\author{Ali Rakhshan and Hossein Pishro-Nik $^{1}$ and Evan Ray $^{2}$
\thanks{*This work was supported by the NSF under CCF 0844725.}
\thanks{$^{1}$Ali Rakhshan and Hossein Pishro-Nik are with the Department of Electrical Engineering,
        University of Massachusetts, Amherst, MA 01003-9305, USA
        {\tt\small \{arakhshan,  pishro\}@ecs.umass.edu}}%
\thanks{$^{2}$Bernard D. Researcheris with School of Mathematics and Statistics, University of Massachusetts,
        Amherst, MA 01003-9305, USA
        {\tt\small Ray@math.umass.edu}}}


%


\maketitle

\maketitle
\thispagestyle{empty}
\pagestyle{empty}

\begin{abstract}
Adapting the functioning of the collision warning systems to the specific drivers' characteristics
is of great benefit to drivers.  For example, by customizing collision warning algorithms we can minimize false alarms, thereby reducing injuries and deaths in highway traffic accidents. In order to take the behaviors of individual drivers into account, the system needs to have a Real-Time estimation of the distribution of brake response times for an individual driver.
In this paper, we propose a method for doing this estimation which is not computationally intensive and can take advantage of the information contained in all data points.
\end{abstract}

\section{Introduction}

Traffic accidents in the United States cause over 30,000 deaths each year \cite{USDoT:TrafficFatalities}.  Some of these accidents could be prevented or reduced in severity if the drivers involved were warned in time to slow or steer to avoid the accident.
 Collision warning systems do reduce the behaviors that lead to crashes
 .
  Radical improvements in the effectiveness of collision warning systems are now possible due to the progress that is being made in Vehicular Ad Hoc Networks (VANET
 ).
  Vehicular ad hoc networks potentially allow all vehicles to communicate with each other (V2V) and with technologies embedded in the road infrastructure (V2I).

The effectiveness of warnings depends on how much time the driver needs to react. Therefore, to be as effective as possible, accident warning systems should be tailored to the specific characteristics of the driver.  An important aspect of that is the distribution of brake response times (BRT) for each particular driver.  The BRT is the time elapsed between a stimulus such as a lead car braking or traffic signal changing color and a braking response by the driver. In this paper we describe a method for estimating the distribution of BRTs for a particular driver using data from a VANET system which has information about the positions, velocities, accelerations of cars on the roads, 
 and the status and position of traffic signals. Then, we will be able to use the estimated distribution to adapt the system to drivers' characteristics \cite{Tunning}. The paper is organized as follows.  In section \ref{section:Related Work} we review the relevant literature formally defining the BRT and related quantities, 
  and outlining some methods that have been proposed to estimate a driver's BRT.
  Section \ref{estimation} outlines the methods that can be used to estimate what the distribution of a driver's BRTs would be if he or she did not intentionally delay braking.  In section \ref{conc} our concluding remarks are discussed.

\section{Related Work}\label{section:Related Work}
\subsection{Basic Ideas: Perception-Reaction Times and Brake Response Times}

\begin{figure}[!t]
	\centering
\framebox{\parbox{3in}{\includegraphics[width=3in]{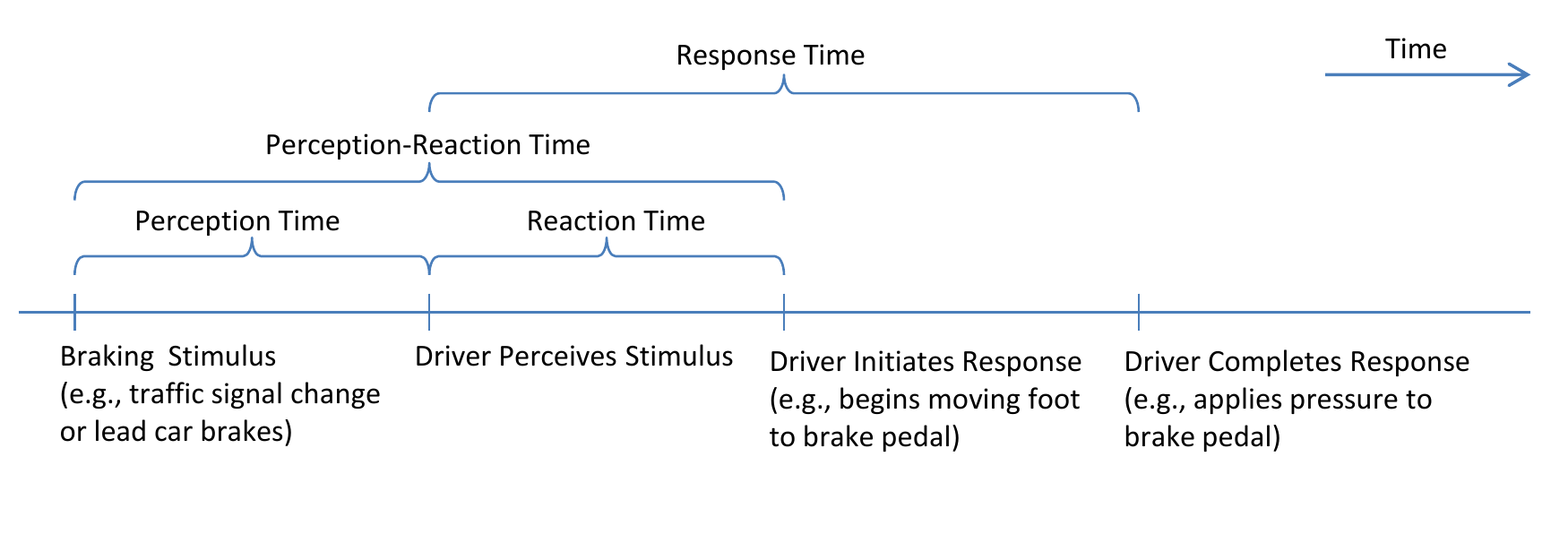}
}}
		
	\caption{The scheme for defining perception reaction times as given by Koppa \cite{Koppa:HumanFactors}.}
	\label{fig:KoppaPRTIllustration}
\end{figure}

The time required to respond to a stimulus can be divided into several distinct phases.  One such division is given by Koppa \cite{Koppa:HumanFactors}.  He defines the perception time as the amount of time it takes for an individual to recognize that an event has occurred.  The reaction time is then the time elapsed from detection of a stimulus to the start of a response.  The response time includes the reaction time as well as the time required to complete the response.  
 These divisions are illustrated in Fig. \ref{fig:KoppaPRTIllustration}.

\begin{figure}[!t]
\centering
\framebox{\parbox{2in}{\includegraphics[width=2in]{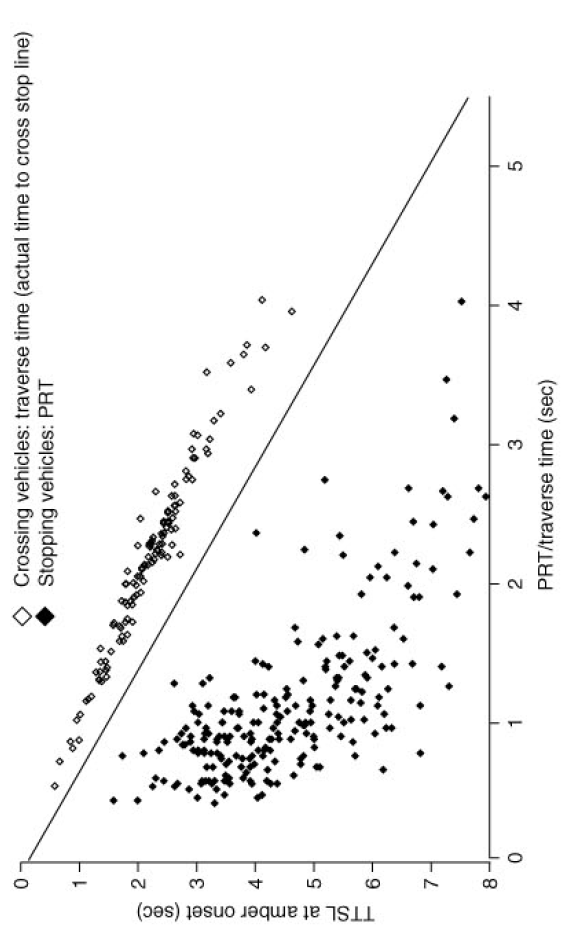}
}}
\caption{Rotated plot from Goh and Wong of observed brake reaction times (PRT in their terminology) vs. time headway to traffic signal \cite{GohWong:DriverPRTDuringSignalChange}.  Points above the diagonal line correspond to cars that did not stop at the intersection.}
\label{fig:GohWongRTPlot}
\end{figure}

There is some ambiguity in this definition of the reaction and response times in that we must specify what is meant by the response.  Commonly in driving studies, the response is operationally defined to be the act of braking
.
 This operational definition is convenient because it is relatively easy to measure when the brakes have been applied.  However, a difficulty with this definition is that a driver may intentionally delay braking, for instance if there is a large space between the driver and a traffic signal or leading car.  This means that measured response times may be larger than the drivers' ``true'' response times \cite{GohWong:DriverPRTDuringSignalChange}.
This delay is illustrated in the data plot reproduced in Fig. \ref{fig:GohWongRTPlot}, which is taken from an article by Goh and Wong \cite{GohWong:DriverPRTDuringSignalChange}.  We have rotated the plot to clarify that we view time headway as the independent variable and response time as the dependent variable.  In this plot the horizontal axis shows the driver's time headway to a traffic signal at the time it turned from green to yellow and the vertical axis shows the measured brake reaction time for drivers who braked (or the actual time to pass the signal for those drivers who ran the light).  We see that when the driver is a larger distance from the traffic signal, their measured brake response time is larger -- likely because they chose to delay braking.

\begin{figure}[!t]
	\centering
\framebox{\parbox{3in}{\includegraphics[width=3in]{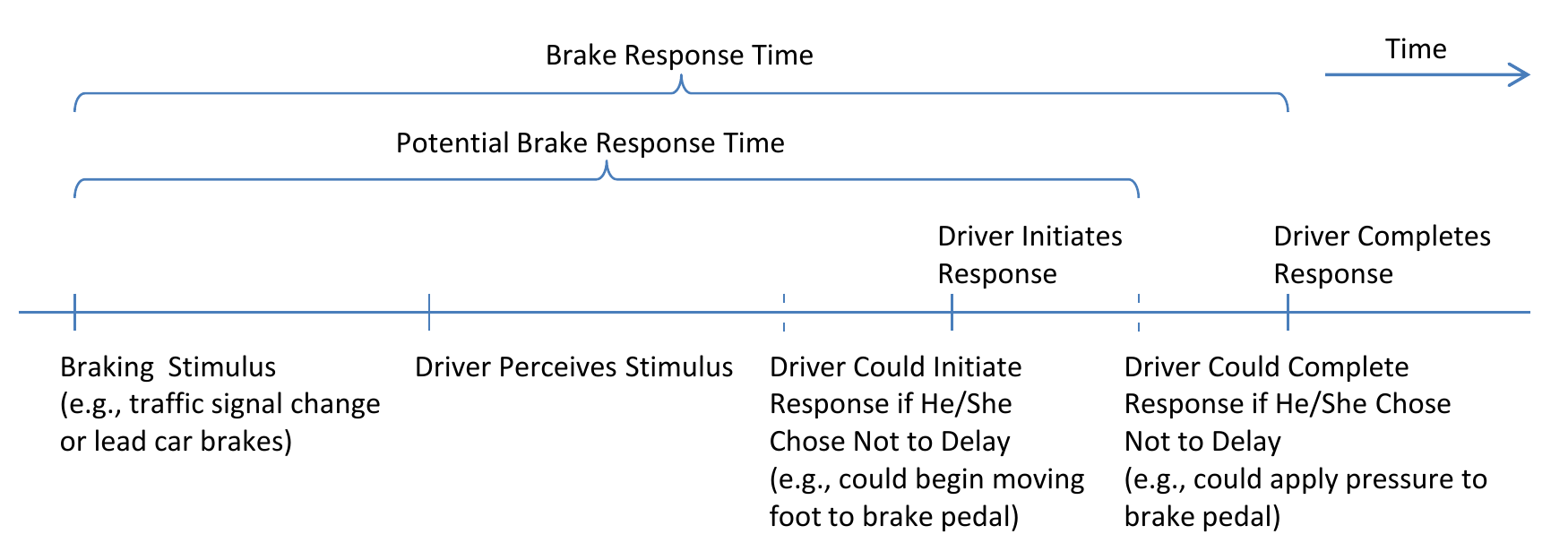}
}}
	\caption{An illustration of the potential brake response time and brake response time.}
	\label{fig:OurPRTIllustration}
\end{figure}

In this paper we define the potential brake response time (PBRT) as the time that a driver could have braked in if he or she did not choose to delay braking, which is the relevant quantity for the purposes of an accident warning system.  We will use the term ``brake response time'' (BRT) to refer to the observed quantity, the time elapsed between a stimulus such as a traffic signal color change and when the driver applies pressure to the brake pedal.  These definitions are illustrated in Fig. \ref{fig:OurPRTIllustration}.  The estimation of BRT and PBRT both present technical difficulties.  We review methods that have been proposed to estimate these quantities by previous researchers in the next two subsections.

\subsection{Estimation of Brake Response Time from Car-Following Data}
\label{litrev:BRTEst}
Several previous studies have examined how BRT can be estimated from car-following data automatically.  Here we review some of these methods, focusing in particular on the effectiveness of these algorithms for obtaining an accurate estimate of the BRTs to several distinct events and on the feasibility of implementing them with the limited computational resources available in an on-board computer system in a car.
The idea proposed by Zhang and Bham \cite{ZhangBham:EstDriverRT} 
 is based on intuitive reasoning about the relationships between the distances, speeds, and accelerations of two cars when the following car reacts to an action taken by the lead car.  The starting point in their algorithm is to identify two cars which go for a period of at least 4 seconds in which they are separated by less than or equal to 250 feet and their speeds are within 5 ft/s, or 1.52 m/s.  These cars are said to be in a steady state.
The advantages of this method are that it is intuitively reasonable, relatively easy to implement, and it yields reasonable reaction time estimates.  However, the requirement that the cars be in steady state is restrictive.
To obtain more information about drivers' reaction times, it would be helpful to extend this approach to estimate reaction times in other situations than the steady state.
Another approach was taken by Ahmed, who specified a reaction time distribution as part of a larger model of car-following behavior, and estimated all parameters of this model jointly through maximum likelihood techniques \cite{Ahmed:ModelingDrivers}.
However, the maximum likelihood estimates had to be obtained numerically, which is computationally intensive due to the complexity of the model.  Therefore, this method would not be practical to implement in a collision warning system where the BRT distribution must be obtained with limited computing resources.

\subsection{Estimation of Potential Brake Response Times from Observed Brake Response Times}
\label{litrev:PRTEst}

Most of the previous studies have addressed the problem of estimating the distribution of ``true'' reaction times based on observed brake response times.  All of these studies examined this problem in the context of traffic signals, and focused on estimation of population distributions, rather than distributions of response times for a particular individual.
Maxwell and Wood simply used the mode response time as a point estimate for the average brake response time in a population, arguing that this measure would be less sensitive to large reaction times that include a delay \cite{MaxwellWood:ReviewTrafficSignal}.  Goh and Wong take a more sophisticated approach \cite{GohWong:DriverPRTDuringSignalChange}.  They define a transitional zone (TZ) based on the time headway between the driver and the traffic signal at the time that it changes to yellow.  This TZ is ``an empirically calibrated range of time headways suitable for identifying drivers with realistic stop-or-cross decisions" \cite{GohWong:DriverPRTDuringSignalChange}.  Essentially, to estimate response times they limit the sample to those cars with a time headway of $\leq$ 4 seconds.  Nearly all cars that chose not to stop at the light were within the 4-second threshold; thus, this threshold includes cars with a ``real'' choice between stopping and continuing on.  This is illustrated in Fig. \ref{fig:GohWongRTPlot} above.
This analysis does suffer from some limitations.  First, by restricting the sample to those cars within the TZ, they lose the information contained in those other data points.  This is a particularly critical problem in our application, where we wish to learn about response times for a particular driver.  We may not have the chance to observe response times very frequently for a single driver; it would therefore be helpful to be able to use all observed data points rather than just those with a time headway of 4 seconds or less.  Second, although the relationship between time headway and BRT is reduced when the sample is restricted to cars with time headway of $\leq$ 4 seconds, a relationship can still be seen in the plot in Fig. \ref{fig:GohWongRTPlot}.  This suggests that some of the measured response times may still include a delay even within the TZ.
In order to be successful in tuning collision warning 
algorithms to individual drivers, we will need a model which provides us with an estimate of the average driver's brake reaction time as well as the individual driver's response time.  The mix of drivers on the road is constantly changing, with new drivers joining and other, usually older, drivers leaving.  Thus when there is no information on an individual, the average response times can be used.  As more information about an individual driver's response times becomes available, the system can switch from the general estimate of brake response time to the individual driver's estimated brake response time.
\section{Estimating the Distribution of Potential Brake Response Times}
\label{estimation}

\subsection{General Discussion}

In this section we discuss the construction of a statistical model for the distribution of brake response times, and how this model can be used to estimate the distribution of potential brake response times for a particular individual.
Virtually every study to examine reaction times has found that the population distribution of reaction times is skewed right and several have shown that it is well approximated by a lognormal distribution 
(\cite{GohWong:DriverPRTDuringSignalChange}, 
\cite{Koppa:HumanFactors}, \cite{MaxwellWood:ReviewTrafficSignal},
  \cite{ZhangBham:EstDriverRT}).  A close examination of the plot in Fig. \ref{fig:GohWongRTPlot} indicates that the distribution of BRTs is also skewed right at a fixed value of time headway.
  It is reasonable to assume that brake reaction times are skewed right within individuals as well.  We therefore adopt a lognormal model for brake reaction times, modelling the logarithm of the observed BRT as normally distributed conditional on the time headway.
This lognormal model also has the advantage of automatically correcting for some differences in the variance of the BRT distribution at different time headways and across individuals.  From the plot in Fig. \ref{fig:GohWongRTPlot}, we can see that as the time headway increases, the mean BRT and the variance of the BRTs both increase.  Similarly, it seems likely that some individuals have lower or higher mean reaction times than other drivers, and that the variance in the BRT distribution varies across individuals as well.  Specifically, it is likely that individuals with a low mean reaction time also have a low variance in their reaction times, whereas individuals with a high mean reaction time also have a high variance in their reaction times.  These differences in the variance of brake reaction times will be approximately corrected by modelling the logarithm of the BRT.
It also seems likely that the mean and variance of the brake response time distribution depend on several other variables.  An important factor that will be accounted for in our model is the stimulus type (e.g. traffic signal vs. lead car decelerates).
 However,
  some of the other factors will not generally be available to the accident warning system, so their effects will be absorbed into the error term of our model.

\subsection{The Model}

\begin{figure}[!t]
	\centering
\framebox{\parbox{1.5in}{\includegraphics[width=1.5in]{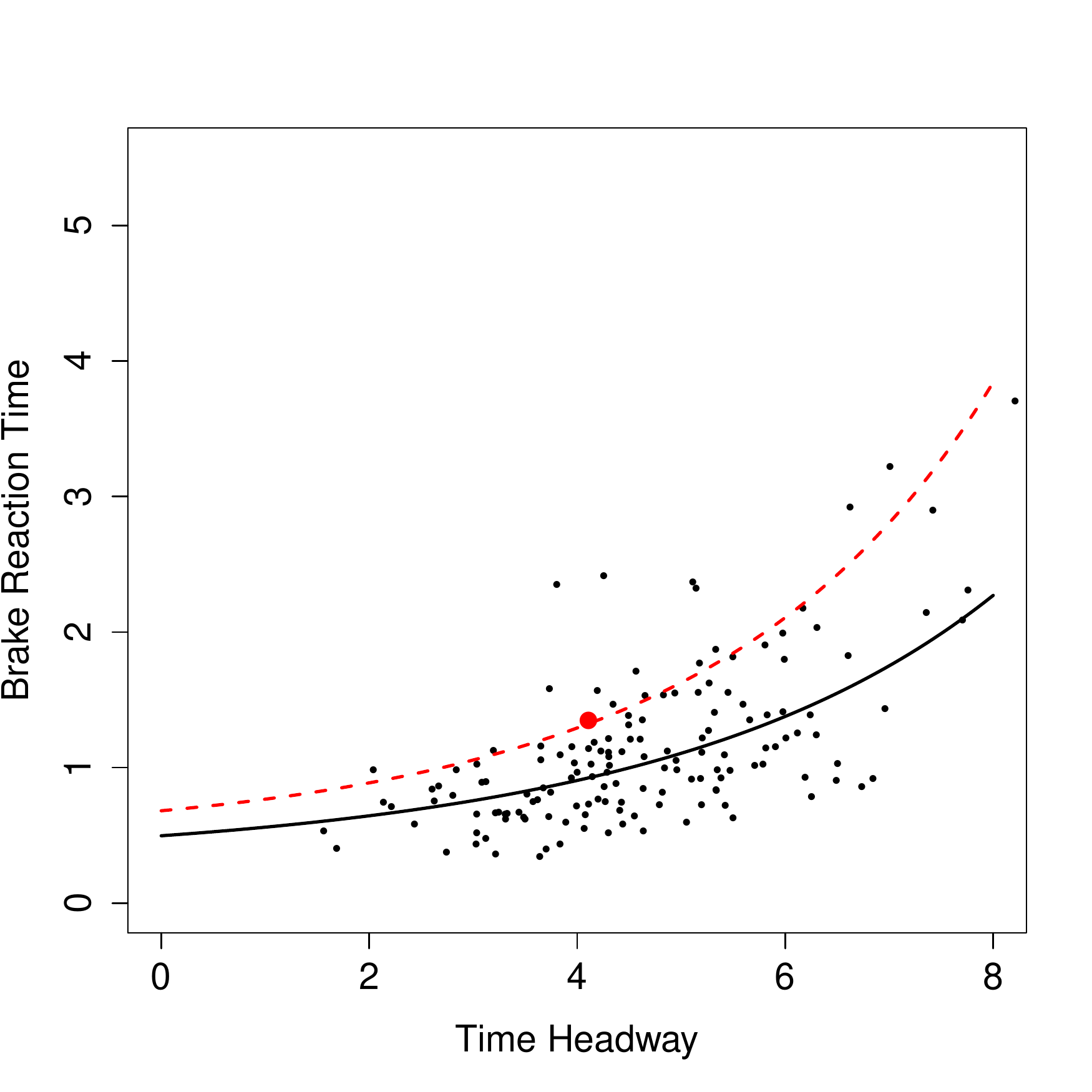}
}}
		\quad
\framebox{\parbox{1.5in}{\includegraphics[width=1.5in]{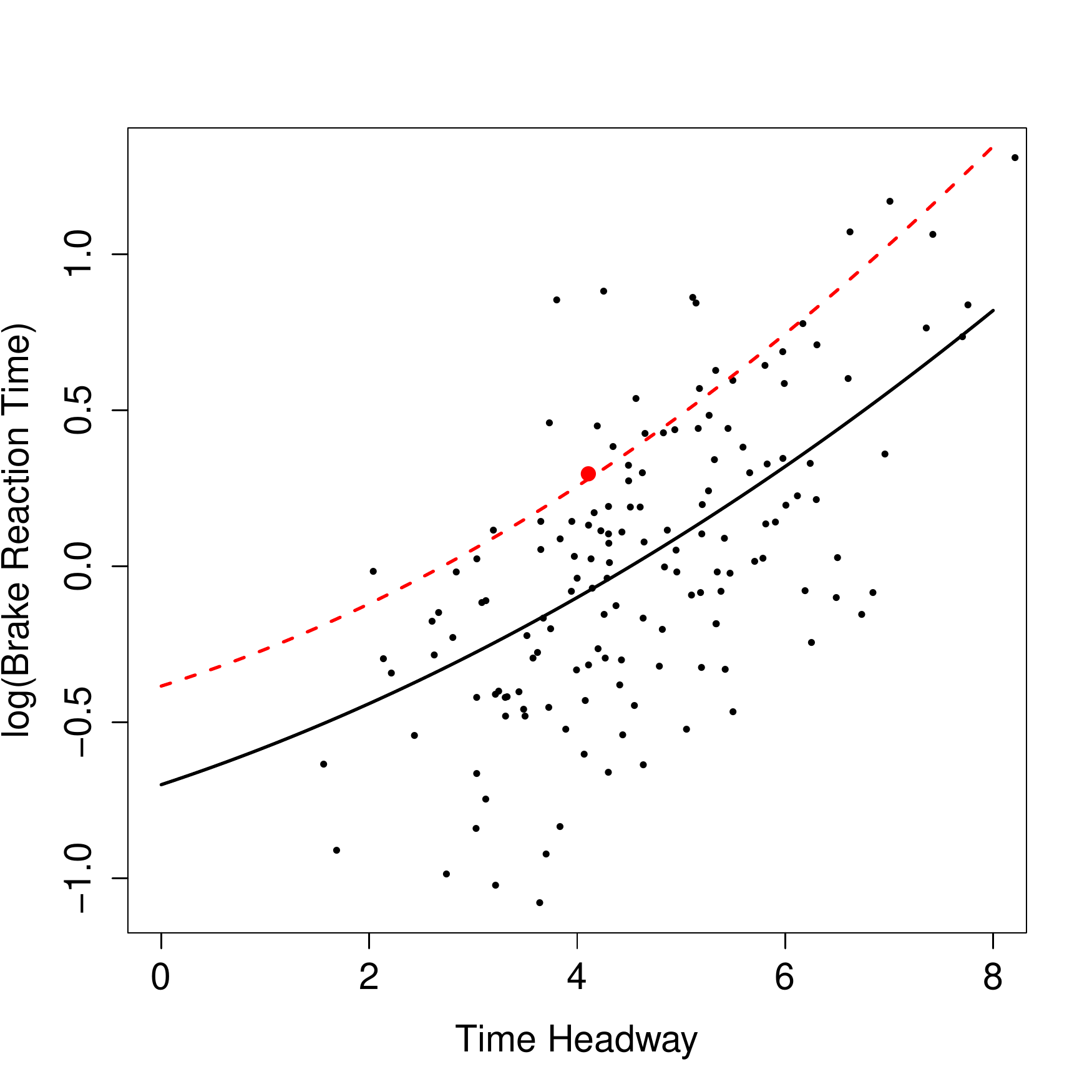}
}}
	\caption{An illustration of the model based on a simulated data set.  All parameters were chosen for the simulation so that the simulated data would be reasonably similar to that in Fig. \ref{fig:GohWongRTPlot}, from Goh and Wong \cite{GohWong:DriverPRTDuringSignalChange}.  Each plot shows simulated data for just one stimulus type.  The black curve represents the population-average relationship between time headway and brake reaction time, $X\beta$.  The red curve represents the relationship between time headway and brake reaction time for one individual, $X(\beta + \gamma)$.  The red point is an observation for that driver.}
	\label{fig:ModelIllustrationSim}
\end{figure}

Using just the time headway as an explanatory variable, the general ideas above can be formalized in the following model:
\begin{align}
\by_{d} &\sim N(X\beta + X\gamma_{d}, \sigma^2 I ) \nonumber \\
\gamma_{d} &\sim N(0, \Sigma_\gamma) \label{ModelStatement}
\end{align}
In this model,
\begin{itemize}
	\item $d$ indexes the driver
	\item $\by_{d}$ is a vector of the logarithms of observed reaction times for a particular driver.
	\item $X$ is a matrix of covariates, detailed further below.
	\item $\beta$ is a fixed vector of unknown coefficients.
	\item $\sigma^2$ is an unknown scalar.
	\item $\gamma_d$ is a random vector of unknown coefficients.
	\item $\Sigma_\gamma$ is an unknown matrix.
\end{itemize}
The basic idea of this model is that, conditional on the time headway, the distribution of BRTs for an individual driver has a mean which is given by an overall population mean, $X\beta$, plus an offset due to the particular characteristics of that driver, $X\gamma_d$.  This is illustrated in Fig. \ref{fig:ModelIllustrationSim}.  It is assumed that the parameters $\gamma_d$ determining the individual's offset to the overall mean follow a multivariate Normal distribution in the population.  This is a linear mixed effects model
 (\cite{McCullochetal:GLMM}, \cite{Searleetal:VC}, \cite{RavishankerDey:LMT}).
A key assumption made in this model specification is that after the log transformation, the covariance matrix $\cov[\by_{d}]$ has the simple form $\sigma^2 I$.  This assumption could fail to hold in a number of ways, but it makes the calculations much easier.
We now consider the form of the mean $X(\beta + \gamma_d)$ in more detail.  From the plot in Fig. \ref{fig:GohWongRTPlot}, we saw that the mean brake reaction time was an increasing function of time headway.  Since the logarithm is a monotonically increasing function, it follows that the logarithm of the BRT is also an increasing function of time headway.  For flexibility, we allow the possibility that the log BRTs are a quadratic function of time headway.  We also allow for the possibility that the relationship between time headway and BRT is slightly different for each of the different stimulus types.  For instance, it could be that drivers have a faster BRT at low time headways and the average BRT increases more rapidly as a function of time headway when the stimulus is a lead car braking than when it is a traffic signal changing to yellow.  These considerations lead to the following possible form of the mean log-BRT as a function of time headway:
\begin{equation}
\nonumber E[y_{dsi}] =
\end{equation}
\begin{equation} \label{quad}
\beta_{s,0} + \beta_{s,1} t_{dsi} + \beta_{s,2} t_{dsi}^2 + \gamma_{d,s,0} + \gamma_{d,s,1} t_{dsi} + \gamma_{d,s,2} t_{dsi}^2
\end{equation}
In equation (\ref{quad}), $d$ indexes the driver, $s$ indexes the stimulus type, and $i$ indexes the observation (so if we have 5 different BRT observations for a particular driver and stimulus type, $i$ will vary from 1 to 5).  As before, $y_{dsi}$ is the log brake reaction time, and $t_{dsi}$ is the time headway at the time of the stimulus.  The subscript $s$ on the $\beta$ and $\gamma$ terms indicate that the values of those coefficients depend upon the stimulus type $s$.  To make this concrete, if this mean function is adopted and there are $S = 3$ different stimulus types under consideration with $n_{ds}$ observations for driver $d$ under stimulus type $s$, $\beta$ and $\gamma_d$ are $9 \times 1$ vectors and the portion of the $X$ matrix corresponding to observations for driver $d$ will be of the following form:
$$\begin{bmatrix} 	\centering
1 & t_{d11} & t_{d11}^2 & 0 & 0 & 0 & 0 & 0 & 0 \\
 1 & t_{d12} & t_{d12}^2 & 0 & 0 & 0 & 0 & 0 & 0 \\
\vdots & \vdots & \vdots & \vdots & \vdots & \vdots & \vdots & \vdots & \vdots \\
 1 & t_{d1n_{d1}} & t_{d1n_{d1}}^2 & 0 & 0 & 0 & 0 & 0 & 0 \\
 0 & 0 & 0 & 1 & t_{d21} & t_{d21}^2 & 0 & 0 & 0 \\
\vdots & \vdots & \vdots & \vdots & \vdots & \vdots & \vdots & \vdots & \vdots \\
 0 & 0 & 0 & 1 & t_{d2n_{d2}} & t_{d2n_{d2}}^2 & 0 & 0 & 0 \\
 0 & 0 & 0 & 0 & 0 & 0 & 1 & t_{d31} & t_{d31}^2 \\
\vdots & \vdots & \vdots & \vdots & \vdots & \vdots & \vdots & \vdots & \vdots \\
 0 & 0 & 0 & 0 & 0 & 0 & 1 & t_{d3n_{d3}} & t_{d3n_{d3}}^2
\end{bmatrix}$$
\subsection{Training the Model: A Fit Using Data from Driving Simulations}
For training the model, we assume data are gathered for $D$ subjects in a driving simulation.  If possible, we prefer to gather data from real drivers on the road, but this is likely to be too difficult to be feasible.  This being the case, we will take precautions to address concerns about using results from a driving simulation to learn about response times for drivers in real life driving situations. The subjects in the study will be a representative sample of the overall population of drivers who will be using the collision warning system.  Brake responses for each subject will be elicited at a variety of levels of expectancy.  To improve the statistical analysis, responses will also be collected at a range of time headways for each stimulus type.  To separate the effects of expectancy and any other variables that may be included in the model, the combinations of these factors will be randomized (for example, we will have some observations where the braking stimulus was more and less surprising at different levels of the time headway variable).
For each driver, we have multiple observations of reaction times for each stimulus type.  These data can be used to estimate the unknown quantities $\beta$, $\sigma^2$, and $\Sigma_\gamma$ in this model using standard statistical techniques implemented in the lmer function of the lme4 library in R.  We will use a subscript of $(tr)$ to indicate quantities obtained from this training data set; in particular, let $X_{(tr)}$ be the covariate matrix obtained using data from this data set and denote the estimates by $\widehat{\beta}_{(tr)}$, $\widehat{\sigma}^2_{(tr)}$, and $\widehat{\Sigma}_{\gamma(tr)}$.  $\widehat{\beta}_{(tr)}$ can be written as $\widehat{\beta}_{(tr)} = (X_{(tr)}' V_{(tr)}^{-1} X_{(tr)} )^{-} X_{(tr)}' V_{(tr)}^{-1} \by_{(tr)}$, where $V_{(tr)} = \cov(\by_{(tr)}) = X_{(tr)} \Sigma_\gamma X_{(tr)}' + \sigma^2 I$ and the superscript $^{''-''}$ denotes a generalized inverse.  The estimates $\widehat{\sigma}^2_{(tr)}$ and $\widehat{\Sigma}_{\gamma(tr)}$ can be found through numerical maximum likelihood techniques.
\subsection{Real Time Estimation of the PBRT Distribution for One Driver}
\label{subsubsec:FreqRealTimeReactionDistEst}
We estimate the distribution of PBRTs for a particular driver in two steps.  First, we establish the relationship between the covariates and BRT for that driver.  Then we use this relationship to estimate the distribution of PBRTs by using values of the covariates at which the BRT does not include an intentional delay to braking.
\subsubsection{Estimating the Relationship Between Time Headway and BRT for One Driver}
As data are gathered in real time for an individual driver $d^*$, our goal is to estimate the driver's offset $\gamma_{d^*}$ to the population-average regression coefficients $\beta$.  This is estimated by the Best Linear Unbiased Predictor (BLUP).
Intuitively, we might expect that if a particular driver has a higher than average brake response time in one stimulus type, they are likely to have a higher than average brake response time in other stimulus types as well.  Similarly, if they are particularly sensitive to the time headway in one situation, they are more likely to be sensitive to the time headway with other stimulus types.  This intuition suggests that the covariance matrix $\Sigma_\gamma$ will have non-zero off-diagonal entries; that is, there is some degree of correlation among the $\gamma_d$ coefficients.  Because of this correlation, observations from one stimulus type can give us information about the coefficients in the other stimulus types.  For example, if we make some observations of driver brake response times in the traffic light setting which give positive estimates of the $\gamma_d$ coefficients for that stimulus, a positive correlation between the coefficients might lead to positive estimates of the coefficients for other stimuli as well.
To reduce the computational complexity of computing the BLUP, we assume that the information about the unknowns $\beta$, $\sigma^2$, and $\Sigma_\gamma$ that is provided by the training data set from the driving simulator is much greater than the information provided by the data from this individual driver.  That is, the estimates $\widehat{\beta}_{(tr)}$, $\widehat{\sigma}^2_{(tr)}$, and $\widehat{\Sigma}_{\gamma(tr)}$ obtained from the training data set above are very similar to what we would obtain if we estimated them using the combined training data set with the observations for this driver.  If this assumption holds, we can approximate the BLUP using the estimates of these quantities found with the training data set, which saves the computational effort of re-fitting the model every time we observe a new reaction time.

Let $X_{d^*}$ be the covariate matrix $X$ as in the full model, but formed using only the data from driver $d^*$.  The BLUP of $\gamma_{d^*}$ is
\begin{equation}
\nonumber \tilde{\gamma}_{d^*} = \Sigma_\gamma X_{d^*}' V_{d^*}^{-1} (\by_{d^*} - X_{d^*} \widehat{\beta})
\end{equation}
where $V_{d^*} = \cov(\by_{d^*}) = X_{d^*} \Sigma_\gamma X_{d^*}' + \sigma^2 I$.  Ordinarily $\widehat{\beta}$ would be estimated from all of the data, but by our assumption above we will instead use the estimate $\widehat{\beta}_{(tr)}$.  The formula for the BLUP still involves the unknowns $\sigma^2$ and $\Sigma_\gamma$.  We estimate the BLUP by plugging in the estimates of these quantities obtained from the training data above.  Denoting this estimated BLUP by $\hat{\gamma}_{d^*}$, we have:
\begin{equation}
\nonumber \hat{\gamma}_{d^*} = \widehat{\Sigma}_{\gamma(tr)} X_{d^*}' \widehat{V}^{-1}_{d^*} (\by_{d^*} - X_{d^*} \widehat{\beta}_{(tr)}),
 \end{equation}
 where
$ \widehat{V}_{d^*} = X_{d^*} \widehat{\Sigma}_{\gamma(tr)} X_{d^*}' + \widehat{\sigma}^2_{(tr)} I.$ The covariance matrix of the BLUP $\tilde{\gamma}_{d^*}$ is given by
\begin{equation}
\nonumber \cov(\tilde{\gamma}_{d^*}) = \cov(\Sigma_\gamma X_{d^*}' V_{d^*}^{-1} (\by_{d^*} - X_{d^*} \widehat{\beta}))
\end{equation}
\begin{equation}
\nonumber = \Sigma_\gamma X_{d^*}' V_{d^*}^{-1} (V_{d^*} - X_{d^*} \cov(\widehat{\beta}_{(tr)}) X_{d^*}') V_{d^*}^{-1} X_{d^*} \Sigma_\gamma
\end{equation}
To estimate the covariance matrix of $\hat{\gamma}_{d^*}$, we plug our approximation to $\widehat{\beta}$, $\widehat{\beta}_{(tr)}$, and our estimates of $\sigma^2$, $\Sigma_\gamma$, and $\cov(\widehat{\beta}_{(tr)})$ into this formula.  Denote this estimated covariance matrix by $\widehat{\Sigma}_{\hat{\gamma}_{d^*}}$.
When no data have been gathered yet, the best predictor is just the vector 0, with covariance matrix $\Sigma_\gamma$. In this case, the estimated mean for the individual is equal to the estimated mean for the population of all drivers.
\subsubsection{Obtaining the Estimated PRBT Distribution}
\begin{figure}[!t]
	\centering
\framebox{\parbox{2in}{\includegraphics[width=2in]{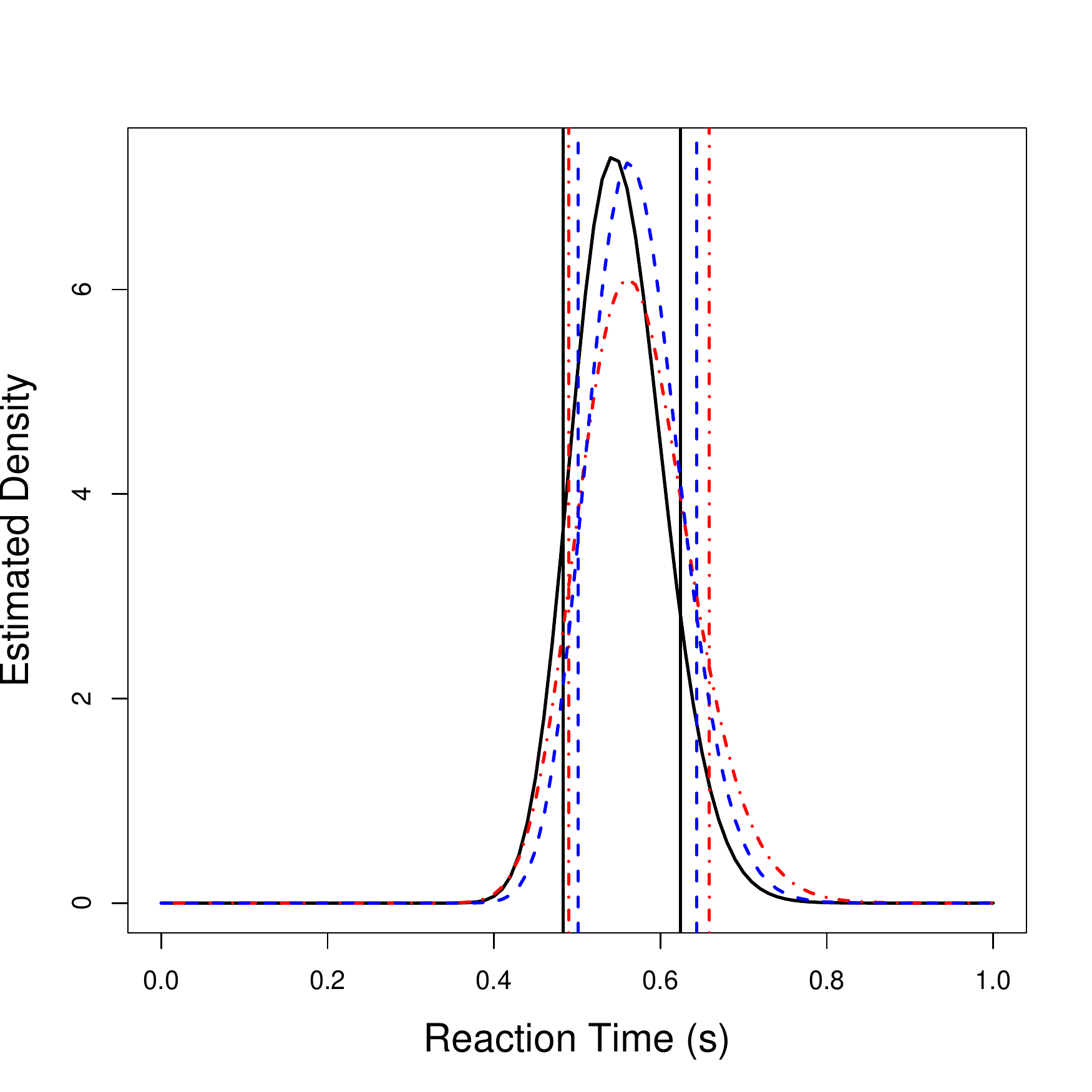}
}}
\caption{Estimates of the distribution of PBRTs for an individual obtained in a simulation.  The black curve represents the individual's ``true'' response time distribution.  The blue curve is the estimated distribution when the variance is taken to be $\hat{\sigma}^2$.  The red curve is the estimated distribution when the variance estimate includes a term for uncertainty in $\widehat{\beta}$ and $\widehat{\gamma_{d^*}}$.  The vertical lines are at the $10^{th}$ and $90^{th}$ percentiles.}
\label{fig:SimRTDistEst}
\end{figure}

The final step is to estimate the distribution of potential brake response times for an individual driver, not including any delays.  For the suggested model form above using a quadratic function of time headway, the intuitive idea is to pick a specific time headway value $t^*$ at which the driver does not have enough time to delay braking, and use that time headway value to evaluate the mean function.  Based on the plots in Fig. \ref{fig:GohWongRTPlot}, it appears that $t^* = 1.5$ might be an appropriate value.  We can then estimate the mean of the driver's log-RTs by plugging $t^* = 1.5$ into the estimated mean function:
$\hat{\mu} = \hat{\beta}_{0} + \hat{\gamma}_{d^*, 0} + t^* (\hat{\beta}_{1} + \hat{\gamma}_{d^*, 1}) + (t^*)^2 (\hat{\beta}_{2} + \hat{\gamma}_{d^*, 2})$.  This provides an estimated mean for the log-reaction time.
There are several options for estimating the variance of the log-PBRT distribution. One simple idea would be to use the estimate $\widehat{\sigma}^2_{(tr)}$ of the quantity $\sigma^2$ in the model statement \ref{ModelStatement}.  However, this does not take into account the uncertainty in our estimate $\hat{\mu}$.  This uncertainty is captured by the prediction error, $(\widehat{\beta}_{(tr)} + \hat{\gamma}_{d^*}) - (\beta + \gamma_{d^*})$.  It can be shown that $\cov((\widehat{\beta}_{(tr)} + \hat{\gamma}_{d^*}) - (\beta + \gamma_{d^*})) = \cov(\widehat{\beta}_{(tr)}) + \cov(\hat{\gamma}_{d^*} - \gamma_{d^*}) - \cov(\widehat{\beta}_{(tr)}, \gamma_{d^*}') - \cov(\gamma_{d^*}, \widehat{\beta}_{(tr)})$, where
\begin{align}
\nonumber\cov(\hat{\gamma}_{d^*} - \gamma_{d^*}) &= \Sigma_{\gamma} - \cov(\hat{\gamma}_{d^*})\\
\nonumber \cov(\hat{\gamma}_{d^*}) &=\\
\nonumber\Sigma_{\gamma} X_{d^*}' ( V^{-1}_{d^*} - V^{-1}_{d^*} X_{d^*} \cov(\widehat{\beta}_{(tr)})&X_{d^*}' V^{-1}_{d^*} )X_{d^*} \Sigma_{\gamma}  
\end{align}
\begin{equation}
\nonumber \cov(\widehat{\beta}_{(tr)}, \gamma_{d^*}') = \cov(\widehat{\beta}_{(tr)}) X_{d^*}' V^{-1}_{d^*} X_{d^*} \Sigma_{\gamma}
\end{equation}
This covariance can be estimated by plugging in estimates of the unknown quantities $V_{d^*}$, $\cov(\widehat{\beta}_{(tr)})$, and $\Sigma_{\gamma}$.  An estimate of the variance of the distribution of log-PBRTs which takes into account our uncertainty about the value of the mean is then
\begin{align}
\nonumber \begin{bmatrix} 1 & t^* & t^{*2} \end{bmatrix} \widehat{\cov}((\widehat{\beta}_{(tr)} + \hat{\gamma}_{d^*}) &- (\beta + \gamma_{d^*}))\begin{bmatrix} 1 & t^* & t^{*2} \end{bmatrix}'\\
\nonumber &+ \hat{\sigma}^2_{(tr)}
\end{align}
When we do not yet have any data, the adjusted variance estimate is
\begin{equation}
\nonumber \begin{bmatrix} 1 & t^* & t^{*2} \end{bmatrix} \widehat{\Sigma}_\gamma \begin{bmatrix} 1 & t^* & t^{*2} \end{bmatrix}'+ \hat{\sigma}^2_{(tr)}.
\end{equation}
The plot in Fig. \ref{fig:SimRTDistEst} shows the resulting distribution estimates obtained in a simulation when these variance estimates are used as the parameters of the distribution of PBRTs.  From this plot we can see that the estimates taking into account uncertainty in the coefficient estimates are more conservative.  On the scale of these simulation results, the difference in the percentiles obtained from these estimates is just a fraction of a second, but the difference could be more significant with real data.  We will use the more conservative value for the estimated variance since it more accurately reflects what we know about the distribution of response times based on the available data.

\begin{figure}[!t]
	\centering
\framebox{\parbox{3in}{\includegraphics[width=3in]{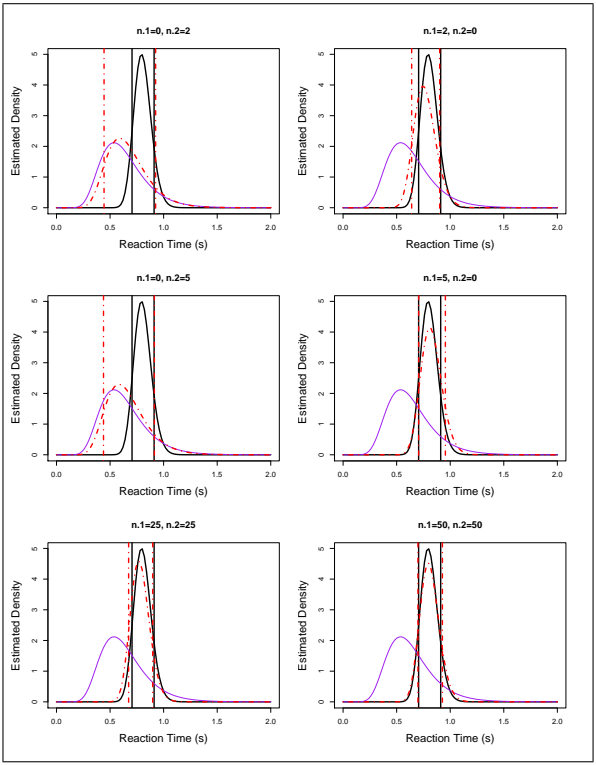}
}}
\caption{Estimates of the distribution of PBRTs for an individual obtained in a simulation with different sample sizes.  The black curve represents the individual's ``true'' response time distribution.  The purple curve represents the distribution of reaction times in the population, which is used as an estimate when the sample size is 0.  The red curve is the estimated distribution.  The vertical lines are at the $10^{th}$ and $90^{th}$ percentiles.}
\label{fig:SimRTDistEstvsSampleSize}
\end{figure}

Fig. \ref{fig:SimRTDistEstvsSampleSize} shows how the estimated reaction time distribution changes with the sample size and the allocation of the sample among the different stimulus types.  These results are dependent upon the parameter values used in the simulation, but they illustrate that observed reaction times for the stimulus type that is used in estimating the PBRT distribution contribute more information than observations in other stimulus types.  This will generally be the case, but our simulation likely shows an extreme example since the correlation among the gamma coefficients for different stimulus types is very low in the simulation.  It could be helpful to run a simulation like this once the training data has been gathered to determine what sample sizes are necessary to get good estimates of the ``true'' PBRT distribution.

We note that computation of the estimated PBRT distribution requires only the operations of matrix inversion and matrix multiplication.  The matrix which must be inverted is $\widehat{V}_{d^*}$, which has dimension $n_{d^*}$, the number of observations for driver $d^*$.  The inversion operation has computational complexity $O(n_{d^*}^3)$.  All of the matrix multiplication operations are between matrices of dimension $9 \times 1$, $9 \times 9$, $9 \times n_{d^*}$, $n_{d^*} \times 1$, or $n_{d^*} \times 1$.  Because multiplying an $n \times m$ matrix by an $m \times k$ matrix has complexity $O(nmk)$, this implies that the complexity of the ``worst'' matrix multiplication operation is $O(9n_{d^*}^2)$ (for the product $X_{d^*}' \widehat{V}_{d^*}^{-1}$).  Therefore the whole computation has complexity $O(n_{d^*}^3)$ when $n_{d^*} > 9$.
\section{Conclusion And Future Work} \label{conc}
In this paper we discussed the need to adapt collision warning systems
to drivers' individual characteristics and proposed a method for doing that
by estimating the distribution of potential brake response times for an
individual driver in real time.  This method uses a statistical model that
was developed based on previously published results about the
population-level brake response times.  However, this model has not yet
been validated using data that includes multiple reaction times for each
driver.  In future work, we will collect this data, fine-tune the model,
and apply it in a collision warning system.
\end{document}